\RequirePackage{fix-cm}
\documentclass[twocolumn]{svjour3}          
\smartqed  
\usepackage{graphicx}
\usepackage{latexsym}
\usepackage[utf8]{inputenc}
\usepackage{multirow}
\usepackage{booktabs}
\usepackage{algorithm,algcompatible}
\algnewcommand\INPUT{\item[\textbf{Input:}]}
\algnewcommand\OUTPUT{\item[\textbf{Output:}]}

\usepackage{pst-node}
\usepackage{tikz-cd} 

\usepackage{amsmath}
\usepackage{amssymb}
\usepackage[colorinlistoftodos]{todonotes}
\usepackage[colorlinks=true, allcolors=blue]{hyperref}
\usepackage{verbatim}
\usepackage{ulem}

\begin{document}
\title{Removable Weak Keys for Discrete Logarithm Based Cryptography
}


\author{Michael John Jacobson, Jr.  \and
       Prabhat Kushwaha$^{*}$
}
\institute{Michael John Jacobson, Jr.  \at
           University of Calgary, Department of Computer Science \\
           2500 University Drive NW, Calgary, Alberta, Canada, T2N 1N4 \\
           \email{jacobs@ucalgary.ca}
           \and
           Prabhat Kushwaha (corresponding author) \at 
           CSE Dept, IIT Kharagpur, West Bengal, India \\
           \email{prabkush@gmail.com} 
           \and {*Both authors contributed equally to all aspects of the paper, and have read and approved the final manuscript.}
}
\date{Received: date / Accepted: date}

\maketitle
\begin{abstract}
We describe a novel type of weak cryptographic private key that can exist in any discrete logarithm based 
public-key cryptosystem set in a group of prime order $p$ where $p-1$ has small divisors. 
Unlike the weak private keys based on \textit{numerical size} (such as smaller private keys, or private keys lying in an interval) that will \textit{always} exist in any DLP cryptosystems, our type of weak private keys occurs purely due to parameter choice of $p$, and hence, can be removed with appropriate value of $p$.
Using the theory of implicit group representations, we present algorithms that can determine whether a key 
is weak, and if so, recover the private key from the corresponding public key.  
We analyze several elliptic curves proposed in the literature and in various standards, giving counts 
of the number of keys that can be broken with relatively small amounts of computation.  Our results show 
that many of these curves, including some from standards, have a considerable number of such weak private keys.
We also use our methods to show that none of the 14 outstanding Certicom Challenge problem instances are 
weak in our sense, up to a certain weakness bound.
\end{abstract}

\section{Introduction}
\label{intro}
Weak cryptographic private keys are those that cause a cryptographic system to have undesirable, insecure behavior.
For example, private keys that can be recovered by an attacker with significantly less 
computational effort than expected can be considered weak.  One recent example of weak keys is described in an April 2019 
whitepaper \cite{ethercombing} by the Independent Security Evaluators, where numerous private keys 
protecting users' Ethereum wallets/accounts were discovered.  Private keys are used to generate corresponding addresses of Ethereum \cite{wood2014ethereum} or Bitcoin \cite{nakamoto2008bitcoin} wallets, and to create digital signatures needed to spend the cryptocurrency. The Ethereum private keys were found easily
because they were very small integers, as opposed to integers of the appropriate bit length.  At the
time of writing this article it is not clear whether Ethereum wallets were assigning these poor keys due to oversight or 
error in the implementation, or whether it was done maliciously.  In any case, the end result is that all the currency in the corresponding accounts was gone.  Note that this type of weak private keys, having small numerical values, \textit{always} exist in all discrete logarithm based cryptosystems, irrespective of the choice of the prime group order $p$.

In this paper, we describe another more subtle type of weak private key that can exist in any discrete logarithm based public-key cryptosystem. These weak keys are special in the sense that they occur purely because of factors of $p-1$, and hence, are \textit{removable} with appropriate choice of $p$, in contrast to always-present smaller private keys. Moreover, our type of weak keys can be quite large in size unlike the small Ethereum keys discussed above, and they can be spread over the whole interval $(1, p)$, and not necessarily in a small sub-interval of $(1, p)$.

As an example, consider the elliptic curve secp256k1 given by
\begin{equation*}
    E: y^2 = x^3 + 7
\end{equation*}
defined over $\mathbb{F}_q$ with $q = 2^{256} - 2^{32} - 2^9 - 2^8 - 2^7 - 2^6 - 2^4 - 1$, and base point
\begin{align*}
    P=      (&5506626302227734366957871889516853432625060\\ 
             &3453777594175500187360389116729240, \\ 
             &3267051002075881697808308513050704318447127\\
             &3380659243275938904335757337482424)
\end{align*}
of prime order 
\begin{align*}
    p =      &115792089237316195423570985008687907852837564\\
             &279074904382605163141518161494337.
\end{align*}
This curve is part of the SEC standard \cite{secg} and is the one used to map users' private keys to Ethereum and Bitcoin public addresses.
The base-$P$ discrete logarithm of the point $Q$
\begin{align*}
     Q = (&1007602026971618930043352141265911168001173\\
          &19792545458764085267675326325395621, \\ 
          &7519344431816503114635930462106279786227214\\
          &2296678797285916994295833810377664)
\end{align*}
is 
\begin{align*}
    \alpha = &64826877121840101682523629462674967702937679\\
             &580369334126295633893540044112329.             
\end{align*}
Although the bit-length of {$\alpha$} is $256$, and thus not weak in the sense of \cite{ethercombing}, given 
only the curve, $P$ and $Q$, we can compute $\alpha$ in less than a second using only $4$ scalar 
multiplications of points on $E$.

Our results are inspired by the work of Maurer and Wolf who showed the equivalence of the discrete logarithm 
problem and the Diffie-Hellman problem in certain cases \cite{maurer1994towards,maurer1999relationship} 
using a technique called implicit group representations. Subsequently, this technique has also been used in \cite{pkkjmc2018}
to estimate a lower bound of the ellipitc curve Diffie-Hellman problem for various standard curves. 
Our work is also closely related to the work of Brown and Gallant \cite{brown2004static} on the static 
Diffie-Hellman problem, which was subsequently rediscovered by and attributed to Cheon \cite{cheon} in the 
context of computing discrete logarithms with auxiliary inputs.  The observation used in all of these works is that 
the discrete logarithm $\alpha$ in a cyclic group $G$ of prime order $p$ can be considered as an element
of the order $p-1$ multiplicative group $\mathbb{F}_p^*$, provided that $\alpha \neq 0$.
Thus, $\alpha = \zeta^i \pmod{p}$ for some integer $0 < i < p-1$, and in principal the discrete 
logarithm can be computed by finding $i$ using a modified version of baby-step giant-step in the order 
$p-1$ group.  Given $d$ dividing $p-1$, either a number of queries to a Diffie-Hellman oracle or 
appropriate auxiliary input can be used to ``lift" the problem to a order $(p-1)/d$ subgroup, where the 
discrete logarithm can be computed more easily.

Kushwaha and Mahalanobis \cite{secrypt17} observed that when $\alpha$ already lies in a sufficiently small 
subgroup of $\mathbb{F}_p^*$, the modified baby-step giant-step 
algorithm of \cite{brown2004static} and \cite{cheon} can be used to find $\alpha$ without any calls to a Diffie-Hellman oracle \cite{Cheon:2016:NAD} or auxiliary input\cite{Cheon2013AGA}.  Our main observation in 
this paper is that, although the approach of \cite{secrypt17} does not appear to result in 
a faster method for computing discrete logarithms in general,
it does reveal a \textit{new type of weak key} 
for discrete logarithm based cryptosystems.  In particular, private keys that can be computed directly 
with the method without any calls to an oracle, provided 
that the subgroup of $\mathbb{F}_p^*$ in which the private key lives is sufficiently small, are \textit{weak}.  

To illustrate the idea further, the underlying reason that the key in the preceding example can be 
computed so easily is that $\alpha$ is in the order $d = {4}$ subgroup of $\mathbb{F}_p^*$; in fact, one 
finds that $\alpha = \zeta_d^{{3}} \pmod{p}$, where $\zeta_d = 7^{(p-1)/{4}} \pmod{p}$ is a generator of the 
order 4 subgroup. 
We can find the discrete logarithm of $\alpha$ to the base $\zeta_d$ using the 
modified baby-step giant-step method in $O((\log p) \sqrt{d})$ group operations, significantly fewer than what is required to compute the discrete logarithm without these considerations.

The main power of our methods thus
occurs when the secret key lies in a small subgroup of $\mathbb{F}^*_p$, allowing one to \textit{detect} whether a 
given private key is weak.  In most cases the probability that a randomly-selected key is weak in this 
sense is very low.  However, a real concern is that a
malicious party could cause users to be assigned weak keys, for example, via hacked or deliberately 
constructed key generation software such as an Ethereum wallet, or hard-coded system parameters 
such as in the Dual\_EC pseudo-random number generator.  The malicious party, knowing that these 
users have weak keys, would be able to recover the private keys at will, as is speculated to have 
occurred in the Ethereum weak keys discovered by the Independent Security Evaluators \cite{ethercombing}.  To further illustrate the threat, we have independently found that there are $343$ Ethereum public addresses and $33$ Bitcoin addresses having private keys between $1$ and $1000$, even though the chances of such occurrences are negligible given that the private key can be any number between $1$ and $2^{256}$.
A similar situation occurred in the well-documented backdoor that was placed in the Dual\_EC 
pseudo-random number generator, which researchers discovered (see \cite{rump_dual_ec}) was enabled in 
part by specifying 
elliptic curve points $P$ and $Q$ where the discrete logarithm of $Q$ to the base $P$ serves as 
trapdoor information for an adversary. Detecting such weak keys is especially important in 
cryptocurrency applications, as well as other applications where solving a single instance of the 
discrete logarithm problem compromises the entire system, such as the Dual\_EC standard and various 
types of identity-based encryption and, more generally, functional encryption.  

Coming back to the previous example, notice that $\alpha$, lying in the subgroup of order $4$, is indeed the private key of a Bitcoin wallet. In fact, there are three active Bitcoin addresses and two Ethereum addresses having private keys in subgroups of size $4$ with multiple transactions to those addresses, most occurring within a few months of writing this article.
These keys are listed in Tables~\ref{tab:bitcoin} and \ref{tab:ethereum}, where $\zeta_4$ denotes a generator of the subgroup of order $4$ and the keys themselves are given by
\begin{align*}
{\zeta_4}^2 = p-1 = \, &11579208923731619542357098500868790\\
&78528375642790749043826051631415181\\&61494336 \pmod p,
\end{align*}
\begin{align*}
{\zeta_4}^3 = \alpha = \, &64826877121840101682523629462674967702\\&93767958036933412629563389354004411232\\
&9 \pmod p, \text{ and}
\end{align*}
${\zeta_4}^4 = 1$. Like the accounts discussed in \cite{ethercombing}, all these \textit{weak} Bitcoin and Ethereum accounts were also empty as of the time this article was written.  Although the keys $\zeta_4^2 = p-1$ (equal to $-1 \pmod{p}$) and $\zeta_4^4 = 1$ fall into the category of keys with small numerical value, the key $\zeta_4^3$ certainly \textit{does not}.

\begin{table*}[htb]
\caption{Bitcoin Addresses with private keys lying in the subgroup of order $d=4$}
\label{tab:bitcoin}       
\begin{tabular}{llcr}
\hline\noalign{\smallskip}
Private \\ key & Bitcoin address &   No. of Txns & Last Txn Date \\
\noalign{\smallskip}\hline\noalign{\smallskip}
 ${\zeta_4}^2$ & 1GrLCmVQXoyJXaPJQdqssNqwxvha1eUo2E & 4 & 19/01/2017 \\
$ {\zeta_4}^3 $ & 1H1jFxaHFUNT9TrLzeJVhXPyiSLq6UecUy & 3 & 16/10/2019 \\
 ${\zeta_4}^4$ &1BgGZ9tcN4rm9KBzDn7KprQz87SZ26SAMH & 40 & 19/12/2019 \\
\noalign{\smallskip}\hline
\end{tabular}
\end{table*}

\begin{table*}[htb]
\caption{Ethereum Addresses with private keys lying in the subgroup of order $d=4$}
\label{tab:ethereum}       
\begin{tabular}{llcr}
\hline\noalign{\smallskip}
Private \\ key & Ethereum address &   No. of Txns & Last Txn Date \\
\noalign{\smallskip}\hline\noalign{\smallskip}
 ${\zeta_4}^2$ & 80c0dbf239224071c59dd8970ab9d542e3414ab2 & 22  & 15/02/2020 \\
 ${\zeta_4}^4$ & 7e5f4552091a69125d5dfcb7b8c2659029395bdf & 713 & 07/02/2020 \\
\noalign{\smallskip}\hline
\end{tabular}
\end{table*}

It is highly unlikely that these keys were generated randomly, demonstrating that relying solely on probabilistic arguments to protect users is not always sufficient. 
A conservative approach would be to eliminate this type of weak key altogether by restricting to groups whose order is a safe prime.  Failing that, it is fortunately
a simple matter to detect weak private keys by computing their multiplicative order modulo $p$; to ensure no weakness whatsoever, one can demand that this order be equal to 
$p-1$.  Furthermore, once can eliminate weak keys like $\alpha$ altogether by restricting to groups whose group order is a safe prime.

It is also possible to test whether a given public key was generated from a weak private key by 
applying the ideas from \cite{secrypt17}, based on \cite{brown2004static} and \cite{cheon}.
This is important because Ethereum or Bitcoin accounts are involved in more than one transaction (see Table~\ref{tab:bitcoin} and~\ref{tab:ethereum}), and 
if attackers discover the private key from the public key given in an Ethereum or Bitcoin transaction, they would be able to spend the cryptocurrency as if they were the legitimate owner of the account.
Our first contribution is therefore to present two algorithms for this task. The first is a baby-step giant-step algorithm that, on input a 
discrete logarithm instance, will determine whether or not the public key is weak, and output the discrete 
logarithm if it is.  The second is a probabilistic algorithm based on the Pollard kangaroo algorithm, that 
will solve the discrete logarithm problem with high probability if the key is in fact weak, 
but may fail to terminate otherwise.
We also present a strategy to verify that a public key is not weak with respect 
to some bound, i.e.\ that the associated private key does not lie in a subgroup of $\mathbb{F}_p^*$ of order less than the bound.  
As an application of our methods, we show that none of the solutions to the 14 outstanding Certicom Challenge problem instances \cite{certicom2009challenge} are in a subgroup of order less than $2^{48}$.

The number of weak keys existing in a particular prime order group can be limited by insisting that the 
group order $p$ be a safe prime ($p-1$ equals twice a prime), or at least has only few small prime 
divisors.  Our second contribution is an analysis of many elliptic curve groups proposed for 
applications and standards.  We observe that the majority of these curves use prime order groups 
for which many weak keys exist.  For example, secp256k1 has more than $2^{24}$ weak keys lying in
subgroups of order less than $2^{32}$ and more than $2^{147}$ weak keys lying in subgroups of order
less than $2^{160}$, which can be computed in roughly $2^{16}$ and $2^{80}$ scalar multiplications,
respectively.  The first type can be computed trivially given the public key, and the latter are on
the threshold of what is likely possible for organizations with sufficient computational resources.

Our paper is organized as follows.  In the next section, we recall the idea of implicit group 
representations, and describe our baby-step giant and kangaroo algorithms to test whether a given
public key comes from a weak private key.  In Section~\ref{sec:analysis} we present our analysis of
elliptic curves proposed for practical applications in terms of the number of weak keys they admit, and 
in Section~\ref{sec:certicom} we present data on experiments using our methods to verify that the 
private keys from the Certicom Challenge elliptic curve discrete logarithm problem instances do not lie 
in subgroups of $\mathbb{F}_p^*$ of order less than $2^{48}$.

\section{Algorithms for Testing Whether a Key is Weak}
\label{sec:2}

In the following, let $G$ be a cyclic group of prime order $p$ generated by an element $g$.  Given another 
element $g_1 \in G,$ the \textit{discrete logarithm problem} is to compute the positive integer $\alpha$ 
with $0 < \alpha < p$ such that $g_1 = g^\alpha$.  

Our algorithms are inspired by the idea of implicit group representations from Mauer and Wolf 
\cite{maurer1994towards} \cite{maurer1999relationship}, in which they were used to prove the equivalence of the discrete logarithm and 
Diffie-Hellman problems in some cases.  They are also closely related to the work of Brown and Gallant 
\cite{brown2004static} on the static Diffie-Hellman problem and Cheon's reformulation 
\cite{cheon,cheon-JCrypt} as the discrete logarithm problem with auxiliary inputs.

The main idea behind all of these works is that $\alpha$, an integer modulo $p$, can also be considered as 
an element of the multiplicative group of a finite field $\mathbb{F}_p^*$, a cyclic group of order $p-1$.  
Let $\zeta$ be a generator of $\mathbb{F}_p^*$.  Then $\alpha = \zeta^i \pmod{p}$ for some integer $i$ 
such that $0 < i < p-1$, and we can thus solve the discrete logarithm problem if we can compute $i$.  When 
trying to solve the discrete logarithm problem we of course do not have to access $\alpha$ itself, rather, 
we have $g_1 = g^\alpha \in G$.  However, the observation that exponentiating elements in $G$ causes 
multiplication in the exponent, i.e.
\begin{equation*}
    g_1^a = (g^\alpha)^a = g^{a \alpha},
\end{equation*}
means that we can implicitly perform the group operation in $\mathbb{F}_p^*$ by exponentiation in $G$.  We 
can also implicitly test for equality in $\mathbb{F}_p^*$ using the fact that $g_a = g^a$ and $g_b = g^b$ 
are equal if and only if $a \equiv b \pmod{p}$.

As a simple example, we can find $\alpha$ by computing $\zeta, \zeta^2, \dots, \zeta^i$ until $\zeta^i 
\equiv \alpha \pmod{p-1}$ by performing these computations using implicit representations.  Thus, we 
compute
\begin{equation*}
    g^\zeta, (g^\zeta)^\zeta = g^{\zeta^2}, \dots, (g^{\zeta^{i-1}})^\zeta = g^{\zeta^i}
\end{equation*}
in the group $G$ via successive exponentiations by $\zeta$ until we have $g^{\zeta^i} = g_1 = g^\alpha$, 
and thus $\alpha \equiv \zeta^i \pmod{p}$ is the solution to the discrete logarithm problem.

Brown and Gallant \cite{brown2004static} and Cheon \cite{cheon,cheon-JCrypt} both observed that this idea 
can be improved given a divisor $d$ of $p-1$.  Then, $\zeta_d = \zeta^{(p-1)/d}$ generates the order $d$ 
subgroup of $\mathbb{F}_p^*$, and $\alpha_d = \alpha^{(p-1)/d}$ lies in this subgroup.  If we had 
$g^{\alpha_d}$ then we could use the algorithm described above to compute $\alpha \pmod{d}$ and, by 
repeating with other divisors of $p-1$ ultimately recover $\alpha$.  Unfortunately we cannot compute 
$g^{\alpha_d}$ from $g$ and $g^\alpha$ using implicit representations, because we would need to 
exponentiate $g^\alpha$ repeatedly by $\alpha$, which we of course do not have.  This is exactly 
where the contributions of Brown and Gallant \cite{brown2004static} and Cheon \cite{cheon,cheon-JCrypt} 
come in.  Brown and Gallant assume that any group element can be raised to the power $\alpha$ via calls to 
a Diffie-Hellman oracle, and Cheon assumes that $g^{\alpha_d}$ is given as auxiliary input to the discrete 
logarithm problem.  Both then compute $\alpha \pmod{d}$ using one application of baby-step giant-step, and 
recover the rest of $\alpha$ with a second iteration of baby-step giant-step.  

However, if $\alpha$ itself happens to lie in the order $d$ subgroup $D$, then we can use the above method to find $\alpha$ by 
computing $g^{\zeta_d}, g^{\zeta_d^2}, \dots$ until we have $g^{ {\zeta_d}^i} = g^\alpha$, and thus $\alpha 
\equiv \zeta_d^i \pmod{p}$.  The difference here is that after at most $d$ iterations we will have either 
found $\alpha$ or verified that $\alpha$ is not in the order $d$ subgroup.  Thus, if $d$ is sufficiently 
small, the discrete logarithm problem can be solved much easier than would otherwise be expected when 
$\alpha$ is in the order $d$ subgroup; any public key for which the corresponding private key has this 
property is thus deemed to be weak.

\subsection{Implicit Baby-Step Giant-Step for Weak Keys}
\label{sec:2.1}

This method can be improved via a direct application of baby-step giant-step to reduce the expected number 
of implicit group operations (exponentiations in $G$) from $O(d)$ to $O(\sqrt{d})$, as described by 
Kushwaha and Mahalanobis in \cite[Theorem 1]{secrypt17}.  We summarize the algorithm here.  Suppose that 
$\alpha = \zeta_d^i$, so that $\alpha$ is in an order $d$ subgroup $D$ of $\mathbb{F}_p^*$.  Then, as in 
standard baby-step giant-step applied to this order $d$ group, there exist unique integers $u$ and $v$ 
such that $0 \leq u,v < m$ and $i = v m -u$ with $m = \lceil \sqrt{d} \rceil$.  We first compute a set of 
baby steps in $G$
\begin{equation*}
    g_1^{\zeta_d^u} = g^{\alpha \zeta_d^u}
\end{equation*}
for $0 \leq u \leq m$ via successive exponentiation by $\zeta_d$.  We next iteratively compute giant steps
\begin{equation*}
    g^{(\zeta_d^m)^v} = g^{\zeta_d^{vm}}
\end{equation*}
for $v=0,1,\dots$ via successive exponentiations by $\zeta_d^m$.  As soon as we find $v$ such that the 
giant step $g^{\zeta_d^{vm}}$ equals the baby step $g^{\alpha \zeta_d^u}$, we have
\begin{equation*}
    \zeta_d^{vm} \equiv \alpha \zeta_d^u \pmod{p}
\end{equation*}
and thus
\begin{equation*}
    \alpha \equiv \zeta_d^i \pmod{p} \text{ with }i \equiv vm - u \pmod{d}
\end{equation*}
is a solution to the discrete logarithm problem.  On the other 
hand, if we compute $m$ giant steps without finding a match, then we conclude that $\alpha$ is not in the 
order $d$ subgroup.  The cost in the worst case is $2m$ exponentiations in $G$, or 
$O((\log p) \sqrt{d})$ group operations in $G$.

Note that our algorithm can be considered as the first phase of Cheon's attack using baby-step giant-step,
for example, as presented in \cite{ITY2011-jip}.  The first difference is that since our purpose is to 
test whether $\alpha$ is in the order $d$ subgroup of $\mathbb{F}_p^*$, only a single application of 
baby-step giant-step is required to either compute $\alpha$ (and not just $\alpha \bmod d$) or verify that 
it is not in the subgroup  The second difference, again because our purpose is to test whether $\alpha$ 
is in the order $d$ subgroup, is that no calls to a Diffie-Hellman oracle nor auxiliary inputs are 
required to obtain group elements of the form $g^{\alpha^j}$.

Various implementations of Cheon's algorithm and numerical results have been reported, including 
\cite{ITY2011-jip,STTY2011-tinyTate,SHITY2012-160bit}.  One important practical improvement that is also 
applicable to our setting is the KKM method \cite{KKM2007-remarks}, due to Kozaki, Kutsuma, and Matsuo.  
The observation is that both the baby steps and giant steps can be written in such a way that each step is 
computed via an exponentiation with the same base element, $g_1$ for the baby steps and $g^{\zeta_d^m}$ 
for the giant steps.  As a result, precomputation tables can be constructed for both phases in such a way 
that each exponentiation is replaced by a constant number of group operations.  Specifically, assuming 
that the base element is $g_t \in G,$ we select an integer $c$, compute $b = \lceil p^{1/c} \rceil$, and 
construct the $c \times b$ dimensional table $T = \{ t_{i,j} \}$ where
\begin{equation*}
    t_{i,j} = g_t^{j b^i} .
\end{equation*}
Then, to compute $g_t^\delta$, we write $\delta$ in base-$b$ as
\begin{equation*}
    \delta = \delta_0 + \delta_1 b + \delta_2 b^2 + \dots + \delta_{c-1} b^{c-1}
\end{equation*}
and compute
\begin{equation*}
    g_t^\delta = (t_{0,\delta_0}) (t_{1,\delta_1}) \dots (t_{c-1,\delta_{c-1}})
\end{equation*}
using only $c-1$ group operations instead of the $O(\log p)$ required for scalar multiplication.  The 
look-up table requires the storage of $cb$ group elements, and Kozaki et.\ al.\ show that the cost to 
compute the table is $O(c p^{1/c})$ group operations.  The total cost of our algorithm using the KKM 
improvement is thus $O(c (p^{1/c} + \sqrt{d}))$ group operations, which is $O(\sqrt{d})$ as long as 
$c \geq 2 \log p / \log d$.  In practice, one chooses an optimal value of $c$ that minimizes the total
number of group operations for the entire algorithm; we will describe our strategy in Section~\ref{sec:certicom}.

\subsection{Implicit Kangaroo Algorithm for Weak Keys}
\label{sec:2.2}

Cheon \cite{cheon} also describes a low-memory variant of his algorithm to solve the discrete 
logarithm problem with auxiliary input, where the two applications of baby-step giant-step are replaced 
with the Pollard kangaroo algorithm.  We describe a specialization of this method to our application of 
determining whether a key is weak.  As with the baby-step giant-step algorithm, this amounts to simply 
running the first kangaroo phase of Cheon's algorithm without using any auxiliary inputs.

As above, we suppose that the discrete logarithm $\alpha = \zeta_d^i \pmod{p}$, so that $\alpha$ is in the order 
$d$ subgroup $D$ of $\mathbb{F}_p^*$ generated by $\zeta_d$.  The main idea is to run the usual kangaroo 
algorithm in $D$ implicitly to compute $i$, using a pseudo-random walk $F:D \rightarrow D$.  The kangaroo 
algorithm would start a wild kangaroo with $\alpha$ and a tame kangaroo with $\zeta_d^{\lceil d \rceil/2}$, and when these two 
random walks collide, compute $i$.  We cannot do this directly, as we only have access to $\alpha$ 
implicitly as $g_1 = g^\alpha$.  However, each of the kangaroo jumps can be mapped to a unique element in 
$G$ via the one-to-one mapping $\phi: D \rightarrow G$, $\beta \mapsto \phi(\beta) = g^\beta$, inducing a 
pseudo-random walk $\bar{F}: G \rightarrow G$ on $G$.   Thus, we start an explicit wild kangaroo at $g_1 = 
g^\alpha$ and an explicit tame kangaroo at $g^{\zeta_d^{\lceil d \rceil/2}}$, and a collision between the two explicit
kangaroos happens exactly when the implicit kangaroos would collide, allowing us to compute $i$ and to 
recover $\alpha = \zeta_d^i \pmod{p}$.  Collisions can be detected using the distinguished points method 
on $G$.

The pseudo-random walk $F$ depends on a pseudo-random function $f:G\rightarrow \{1, 2,..., L\}$ that 
partitions the group $G$ into $L$ partitions of almost equal size (typical values of $L$ are $256, 1024,
2048$) . For the $k^{th}$ partition, there is a small jump $s_k$ such that $1 \leq s_k \leq \sqrt{d}/2$  
subject to the condition that mean step size $m=(\sum_{k=1}^{L} s_k) / L$ is close to $\sqrt{d}/2$, as 
this is required to minimize the overall running time.  Given $f,$ the implicit pseudo-random walk on $D$ 
is defined as
\begin{equation*}
   F: D \rightarrow D; \hspace{0.5 cm} x \mapsto x \cdot {\zeta_d}^{s_{f(g^{x})}},
\end{equation*}
inducing the explicit pseudo-random walk $\bar{F}$ on $G$ defined as
\begin{equation*}
   \bar{F}: G \rightarrow G; \hspace{0.5 cm} g^x \mapsto g^{x \cdot {\zeta_d}^{s_{f(g^{x})}}}.
\end{equation*}
Assuming $F$ is constructed with these properties, Cheon's analysis \cite{cheon-JCrypt} specializes to our 
case and yields an expected running time of $O(\sqrt{d} + \theta^{-1})$ exponentiations in $G$, where 
$\theta$ is the proportion of distinguished points used out of the group $G$.  The number of group 
elements stored (for the distinguished points) is $O(\theta \sqrt{d})$, so $\theta$ may be selected to 
favor either the running time or storage requirement as necessary.

As with our baby-step giant-step algorithm, the KKM optimization \cite{KKM2007-remarks} can be used to 
replace the exponentiations in $G$ with a constant number of group operations at the cost of computing and 
storing a precomputed table.  We describe the details of our implementation in 
Section~\ref{sec:certicom}, including the application of KKM and a precise description of our 
pseudo-random walk and realization of distinguished points.

\subsection{Testing Whether a Key is Weak}\label{sec:weaktest}
We present our approach to test whether the private key corresponding to a given public key is weak 
according to a given bound, i.e. if it belongs to a subgroup of order less than $B$.  This allows one to 
verify that a key is not weak subject to whatever computational bound is feasible.

A simple approach is to run our baby-step giant-step algorithm on all divisors of $p-1$ that are less than 
$B$.  However, this would be inefficient and redundant, because testing whether a key is in a subgroup of 
order $d$ also covers all subgroups of order divisible by $d$.  Thus, we instead generate a list of integers 
$d_1 < d_2 <. ..< d_t \leq B$ dividing $p-1$ such that $d_i \nmid d_j$ for all $1 \leq i < j \leq t$. 
We then apply our baby-step giant-step algorithm to all subgroups of order $d_i$ in the list, thereby 
avoiding redundant computations in subgroups.

For example, for the elliptic curve secp256k1, $p-1$ has 10 divisors bigger than 1 and $\leq 48$, namely 
$2, 3, 4, 6$, $8$, $12, 16, 24, 32, 48$.  In order to test whether a given private key is in 
any of the subgroups of these orders, it suffices to test only the subgroups of orders $32$ and $48$, 
as the first 8 subgroup orders divide $48$, and thus any element of one of these smaller orders is also 
an element of the subgroup of order $48$.

Note that either baby-step giant-step or the kangaroo method described above can be used here if the 
purpose is simply to recover a private key.  However, if the purpose is to verify rigorously that a key 
is \textit{not} weak, i.e.\ does not lie in a subgroup of bounded order, then baby-step giant-step must be 
used.  If baby-step giant-step fails to recover a key by searching in a particular subgroup, we can 
conclude that the key is not in the subgroup.  Due to its probabilistic nature, if the kangaroo method 
fails to recover a key we cannot make the same conclusion.

\section{Assessment of Weak Keys in Recommended and Standard Elliptic Curves}
\label{sec:analysis}
\subsection{Curves Investigated}
We have investigated a number of elliptic curves recommended for practical applications and appearing in 
standards.  The sources of the curves we selected include
\begin{itemize} 
    \item NIST curves \cite{NIST-curves},
    \item Safe curves \cite{safe}, 
    \item Certicom challenge curves \cite{certicom2009challenge}, 
    \item SEC curves \cite{secg},
    \item Brainpool curves \cite{rfc5639}, 
    \item Four$\mathbb{Q}$ curve \cite{FourQ},
    \item the pairing-friendly curves BLS \cite{barreto2002constructing}, BN \cite{barreto2005pairing} and KSS \cite{kachisa2008constructing}, but with updated parameters given in \cite{kiyomura2017secure}, \cite{barbulescu2019updating} and \cite{cfrg2019pairing}.
\end{itemize}

In the tables below, we use the following naming conventions for the various curves considered.  Curve ANSSI refers to the
safe curve ANSSI FRP256v1, and Ed$448$ refers to the safe curve Ed$448$-Goldilocks. The labels of the Brainpool curves 
have also been abbreviated; for example, BP256r1 refers to the Brainpool curve BrainpoolP256r1 and so on.  
Note that we do not include the twisted versions of the Brainpool curves, as they have the same point order as
the non-twisted versions.  

In recent years, there have been tremendous advances in solving the discrete logarithm problem in finite fields. As a result, pairing-based curves over small characteristic fields are no longer safe to use due to the quasi-polynomial attack of Barbulescu et al. \cite{barbulescu2014heuristic}. Moreover, there have also been improvements in the case of medium-sized prime characteristics in a series of papers \cite{joux2006number}, \cite{barbulescu2015tower}, \cite{sarkar2016new}, \cite{kim2016extended}. 
Since most of pairing-friendly curve constructions use medium-sized characteristics fields, previous parameters such as those for BN curves \cite{barreto2005pairing}, BLS curves \cite{barreto2002constructing} and KKS curves \cite{kachisa2008constructing} are no longer applicable to attain the prescribed security level they were defined for.
Therefore, we have examined only the updated parameters as presented in \cite{kiyomura2017secure}, \cite{barbulescu2019updating} and \cite{cfrg2019pairing} for these curves at security level $128$-, $192$-, $256$-bits.

It should also be noted that there are many curves that are part of more than one standard. To ensure each curve 
is mentioned just once in our tables, we follow the label used in the standard which comes first in the list at the 
beginning of this section. For example, the NIST curve K-163 is the same as the Certicom challenge curve ECC2K-238 
and the SEC curve sect163k1, so this curve is listed once in our tables as K-163.
Similarly, for pairing-friendly curves at the $128$-bit security level with updated parameters, BN given in \cite{barbulescu2019updating} and BN-$462$ given in \cite{cfrg2019pairing} are the same; we use the label BN-$462$ in our table.  We also use the label BLS48 in our tables to denote both
BLS48 from \cite{kiyomura2017secure} and BLS48-581 from \cite{cfrg2019pairing} at the 256-bit security level.
The nomenclature for the curve labels used in this paper has been described using Table \ref{tab:curveLabels}. 

 \begin{table*}[htb]
 \caption{Curve nomenclature used for duplicate curves in the tables}
 \label{tab:curveLabels}       
 \begin{tabular}{lllllll}
 \hline\noalign{\smallskip}
 Our Label & NIST\cite{NIST-curves} & Certicom\cite{certicom2009challenge} & SEC\cite{secg} & Kiyomura\cite{kiyomura2017secure} &
 Barbulescu\cite{barbulescu2019updating} & 
 CFRG\cite{cfrg2019pairing} \\
 \noalign{\smallskip}\hline\noalign{\smallskip}
P-192 & P-192 & -& secp192r1 & -& - & -\\
P-224 & P-224 & - &secp224r1 & -& -& -\\
P-256 & P-256 &  &secp256r1 & -& -& - \\
P-384 & P-384 &- & secp384r1 & -& -& - \\
P-521 & P-521 & &secp521r1 &- & -& - \\
K-163 & K-163 & ECC2K-163 & sect163k1 & -& -& - \\
B-163 & B-163 &- &sect163r2 & -& -& - \\
K-233 & K-233 & - & sect233k1 & -& -& - \\
B-233 & B-233 & - & sect233r1 & -& -& -\\
K-283 & K-283 &-&sect283t1 &- &- & -\\
B-283 & B-283 & -&sect283r1 & -& -& -\\
K-409 & K-409 &- &sect409k1 & -& -& -\\
B-409 & B-409 &-& sect409r1 & -& -& -\\
K-571 & K-571 & -&sect571k1 &- &- & -\\
B-571 & B-571 & &sect571r1 & -& -& -\\
ECC2K-238 &- & ECC2K-238 &sect239k1 &- & -& -\\

BN & - & -&- & - & BN & BN-462 \\
BLS48 & - & -&- & BLS48 & - & BLS48-581  \\

\noalign{\smallskip}\hline
\end{tabular}
\end{table*}

\subsection{Analysis of Weak Keys}
Our data are presented in Tables~\ref{tab:weak112}, \ref{tab:weak128}, \ref{tab:weak192}, 
\ref{tab:weak256}, ~\ref{tab:pairing128},~\ref{tab:pairing192} and~\ref{tab:pairing256}.  Table~\ref{tab:weak112} includes all curves providing less than $128$ bits of security, 
where as usual $b$ bits of security indicates that the expected cost to solve the discrete logarithm problem is 
roughly $2^b$.
Table~\ref{tab:weak128} covers security level of approximately $128$ bits but less than $192$, 
Table~\ref{tab:weak192} covers approximately $192$ and up to $256$ bits, and Table~\ref{tab:weak256} covers approximately $256$ bits of security and above. 
Tables~\ref{tab:pairing128},~\ref{tab:pairing192} and~\ref{tab:pairing256} present the data for updated pairing friendly curves for security level 128-, 192- and 256-bits, respectively.

For each curve appearing in the sources listed above, we enumerated the number of weak keys appearing in subgroups of 
size bounded by $B$ for $B=2^{32}, 2^{64}, 2^{128},$ and $2^{160}$.  The cost to determine whether a given
key is weak for each of these bounds is roughly $2^{16},$ $2^{32}$, $2^{64}$, and $2^{80}$ group operations
--- these bounds were selected to give two relatively easy bounds and two at the edge of computations that 
are feasible.  Due to the sizes of numbers occurring in the counts, and in order to 
facilitate an easier comparison, we list the base-2 log (number of bits) of each number as opposed to the 
number itself.  In summary, the data recorded for each curve is as follows:
\begin{itemize}
    \item  Curve label
    \item  $b(p)$:  number of bits of the size of the curve's large prime (subgroup) order
    \item  $b(p_{m})$:  number of bits of the largest prime divisor of $p-1$
    \item  $N_B$:  base-2 log of the number of weak keys with order bounded by $B$.  Since 
    $\phi(d)$ is the number of generators of a cyclic group of order $d$, i.e.\ the number of elements of 
    order exactly equal to $d$, we compute
      \begin{equation*}
        N_B = \log_2 \sum_{\substack{d ~|~ p-1\\ d \leq B}} \phi(d).
      \end{equation*}
    \item  $C_B$:  base-2 log of the worst-case number of group operations required to test whether a key 
    comes from a subgroup of order bounded by $B$ using baby-step giant-step.  Let $R(p,B)$ denote the 
    set of divisors of $p-1$ that must be considered to check whether a key is in a subgroup of order 
    bounded by $B$, i.e.
      \begin{equation*}
          R(p,B) = \{ d_1, \dots, d_t ~:~ d_i \nmid d_j \text{ for all } 1 \leq i < j <= t \}.
      \end{equation*}
    We then compute
      \begin{equation*}
          C_B = \log_2 \sum_{d \in R(p,B)} 2 \lceil \sqrt{d} \rceil
      \end{equation*}
    Note that this cost value measures the worst-case number of scalar multiplications required on the 
    curve.  However, with the KKM method \cite{KKM2007-remarks}, this also measures the worst-case number 
    of point additions required up to a constant factor.
\end{itemize}
The curves are sorted in increasing order of $C_B$ for $B = 2^{160}$, i.e.\ the worst-case cost for determining whether a private key
is in a subgroup of order at most $2^{160}$.

\begin{table*}[htb]
\caption{Curves providing security below $128$ bits }
\label{tab:weak112}       
\begin{tabular}{lrrrrrrrrrr}
\hline\noalign{\smallskip}
Curve & $b(p)$ & $b(p_{m})$ & ${N_{2^{32}}}$ & ${C_{2^{32}}}$ &  ${N_{2^{64}}}$ & ${C_{2^{64}}}$ & ${N_{2^{128}}}$ & ${C_{2^{128}}} $ & ${N_{2^{160}}}$    &  ${C_{2^{160}}} $ \\
\noalign{\smallskip}\hline\noalign{\smallskip}
secp224k1 & 225 & 222 & 2.6 & 2.6 & 2.6 & 2.6 & 2.6 & 2.6 & 2.6 & 2.6 \\
BP224r1 & 224 & 214 & 10.0 & 6.0 & 10.0 & 6.0 & 10.0 & 6.0 & 10.0 & 6.0 \\
secp192k1 & 192 & 167 & 25.4 & 13.7 & 25.4 & 13.7 & 25.4 & 13.7 & 25.4 & 13.7 \\
P-224 & 224 & 196 & 29.0 & 15.5 & 29.0 & 15.5 & 29.0 & 15.5 & 29.0 & 15.5 \\
BP192r1 & 192 & 97 & 11.7 & 6.9 & 11.7 & 6.9 & 108.1 & 55.1 & 108.1 & 55.1 \\
P-192 & 192 & 92 & 17.5 & 9.8 & 17.5 & 9.8 & 109.0 & 55.6 & 109.0 & 55.6 \\
secp112r2 & 110 & 42 & 33.7 & 19.5 & 66.3 & 36.0 & 109.8 & 55.9 & 109.8 & 55.9 \\
sect193r2 & 193 & 109 & 2.0 & 2.0 & 2.0 & 2.0 & 110.2 & 56.1 & 110.2 & 56.1 \\
secp112r1 & 112 & 95 & 17.1 & 9.6 & 17.1 & 9.6 & 111.8 & 56.9 & 111.8 & 56.9 \\
sect113r1 & 113 & 49 & 33.1 & 18.6 & 64.7 & 34.2 & 112.0 & 57.0 & 112.0 & 57.0 \\
sect113r2 & 113 & 39 & 33.7 & 19.0 & 65.5 & 35.2 & 112.0 & 57.0 & 112.0 & 57.0 \\
sect193r1 & 193 & 100 & 22.0 & 12.0 & 22.0 & 12.0 & 121.5 & 61.7 & 121.5 & 61.7 \\
E-222 & 220 & 114 & 8.3 & 5.2 & 8.3 & 5.2 & 121.6 & 61.8 & 121.6 & 61.8 \\
secp128r2 & 126 & 95 & 31.5 & 16.8 & 31.5 & 16.8 & 126.0 & 64.0 & 126.0 & 64.0 \\
secp128r1 & 128 & 57 & 31.8 & 18.0 & 65.9 & 35.4 & 128.0 & 65.0 & 128.0 & 65.0 \\
ECC2K-130 & 130 & 75 & 33.8 & 19.2 & 54.8 & 28.4 & 128.1 & 66.2 & 129.0 & 65.5 \\
sect131r1 & 131 & 116 & 14.5 & 8.3 & 14.5 & 8.3 & 126.3 & 64.4 & 130.0 & 66.0 \\
sect131r2 & 131 & 75 & 30.0 & 16.3 & 55.2 & 28.6 & 104.7 & 53.7 & 130.0 & 66.0 \\
ECC2-131 & 131 & 113 & 17.9 & 9.9 & 17.9 & 9.9 & 126.8 & 65.1 & 130.0 & 66.0 \\
ECCp-131 & 131 & 63 & 19.9 & 11.0 & 65.2 & 34.7 & 128.3 & 66.0 & 130.2 & 66.1 \\
ECCp-191 & 191 & 144 & 1.0 & 2.0 & 47.3 & 24.7 & 47.3 & 24.7 & 144.7 & 73.3 \\
ECC2-191 & 191 & 75 & 10.4 & 6.2 & 50.7 & 26.4 & 124.7 & 63.4 & 149.6 & 75.8 \\
B-163 & 163 & 133 & 29.7 & 15.8 & 29.7 & 15.8 & 29.7 & 15.8 & 156.5 & 79.6 \\
Anomalous & 204 & 71 & 28.2 & 15.5 & 62.9 & 33.3 & 127.2 & 65.5 & 158.1 & 80.2 \\
M-221 & 219 & 88 & 32.5 & 17.7 & 62.9 & 32.4 & 127.8 & 65.2 & 158.4 & 80.3 \\
secp160r2 & 161 & 72 & 21.6 & 11.8 & 21.6 & 11.8 & 93.1 & 47.8 & 159.0 & 80.6 \\
ECCp-163 & 163 & 135 & 28.2 & 15.1 & 28.2 & 15.1 & 28.2 & 15.1 & 159.4 & 80.8 \\
BP160r1 & 160 & 42 & 33.6 & 19.4 & 67.0 & 36.5 & 129.9 & 67.3 & 159.9 & 80.9 \\
secp160r1 & 161 & 105 & 31.5 & 16.9 & 55.3 & 28.6 & 127.7 & 65.4 & 159.2 & 81.1 \\
secp160k1 & 161 & 77 & 18.0 & 10.0 & 18.0 & 10.0 & 94.8 & 48.4 & 159.6 & 81.8 \\
K-163 & 163 & 103 & 19.2 & 10.6 & 59.3 & 30.6 & 121.9 & 62.0 & 160.3 & 82.0 \\
sect163r1 & 162 & 78 & 30.2 & 16.5 & 65.0 & 34.8 & 126.9 & 64.6 & 160.7 & 82.2 \\
ECC2-163 & 163 & 46 & 37.9 & 23.6 & 70.9 & 40.7 & 134.1 & 71.9 & 160.9 & 82.8 \\
\noalign{\smallskip}\hline
\end{tabular}
\end{table*}

\begin{table*}[htb]
\caption{Curves providing security between $128$ and $192$ bits }
\label{tab:weak128}       
\begin{tabular}{lrrrrrrrrrr}
\hline\noalign{\smallskip}
Curve  & $b(p)$ & $b(p_{m})$ & ${N_{2^{32}}}$ & ${C_{2^{32}}}$ &  ${N_{2^{64}}}$ & ${C_{2^{64}}}$ & ${N_{2^{128}}}$ & ${C_{2^{128}}} $ & ${N_{2^{160}}}$    &  ${C_{2^{160}}} $ \\
\noalign{\smallskip}\hline\noalign{\smallskip}
BP256r1 & 256 & 252 & 4.2 & 3.3 & 4.2 & 3.3 & 4.2 & 3.3 & 4.2 & 3.3 \\
BP320r1 & 320 & 278 & 34.7 & 20.4 & 42.4 & 22.2 & 42.4 & 22.2 & 42.4 & 22.2 \\
ECC2-238 & 238 & 182 & 30.2 & 16.4 & 55.6 & 28.8 & 55.6 & 28.8 & 55.6 & 28.8 \\
256KM2 & 257 & 192 & 16.4 & 9.2 & 63.2 & 32.6 & 64.2 & 33.1 & 64.2 & 33.1 \\
ANSSI & 256 & 187 & 33.3 & 18.8 & 64.9 & 34.3 & 69.2 & 35.6 & 69.2 & 35.6 \\
ECCp-239 & 239 & 115 & 22.6 & 12.3 & 22.6 & 12.3 & 128.1 & 65.6 & 136.8 & 69.4 \\
Curve25519 & 253 & 138 & 7.0 & 4.6 & 7.0 & 4.6 & 114.3 & 58.2 & 144.7 & 73.4 \\
secp256k1 & 256 & 109 & 24.1 & 13.1 & 64.7 & 34.2 & 129.4 & 67.0 & 147.9 & 75.0 \\
Four$\mathbb{Q}$ & 246 & 147 & 27.8 & 15.5 & 64.1 & 34.1 & 99.3 & 50.6 & 150.3 & 76.1 \\
K-233 & 232 & 158 & 33.0 & 18.6 & 64.0 & 33.5 & 73.2 & 37.6 & 159.8 & 81.5 \\
B-233 & 233 & 145 & 33.0 & 18.4 & 64.7 & 34.3 & 87.6 & 44.8 & 160.7 & 81.9 \\
B-283 & 282 & 90 & 29.5 & 15.7 & 64.2 & 33.6 & 128.9 & 66.8 & 161.7 & 83.4 \\


sect239k1 & 238 & 104 & 33.0 & 18.8 & 66.5 & 36.0 & 129.0 & 67.0 & 162.1 & 83.7 \\
Curve1174 & 249 & 60 & 35.2 & 21.0 & 68.7 & 38.5 & 133.3 & 71.3 & 164.7 & 86.6 \\
P-256 & 256 & 92 & 36.0 & 21.5 & 69.3 & 38.8 & 133.2 & 70.8 & 165.3 & 86.9 \\
K-283 & 281 & 137 & 38.1 & 23.8 & 71.1 & 40.8 & 133.1 & 70.9 & 165.7 & 87.4 \\
\noalign{\smallskip}\hline
\end{tabular}
\end{table*}

\begin{table*}[htb]
\caption{Curves providing security between $192$ and $256$ bits}
\label{tab:weak192}       
\begin{tabular}{lrrrrrrrrrr}
\hline\noalign{\smallskip}
Curve & $b(p)$ & $b(p_{m})$ & ${N_{2^{32}}}$ & ${C_{2^{32}}}$ &  ${N_{2^{64}}}$ & ${C_{2^{64}}}$ & ${N_{2^{128}}}$ & ${C_{2^{128}}} $ & ${N_{2^{160}}}$    &  ${C_{2^{160}}} $ \\
\noalign{\smallskip}\hline\noalign{\smallskip}
ECCp-359 & 359 & 354 & 5.2 & 3.6 & 5.2 & 3.6 & 5.2 & 3.6 & 5.2 & 3.6 \\
M-383 & 381 & 354 & 26.6 & 14.3 & 26.6 & 14.3 & 26.6 & 14.3 & 26.6 & 14.3 \\
Curve41417 & 411 & 352 & 32.5 & 18.1 & 59.4 & 30.7 & 59.4 & 30.7 & 59.4 & 30.7 \\
Curve383187 & 381 & 299 & 32.0 & 18.1 & 65.5 & 35.3 & 81.4 & 41.7 & 81.4 & 41.7 \\
P-384 & 384 & 281 & 13.5 & 7.8 & 13.5 & 7.8 & 103.3 & 52.7 & 103.3 & 52.7 \\
K-409 & 407 & 299 & 37.2 & 22.9 & 69.0 & 38.8 & 108.8 & 55.4 & 108.8 & 55.4 \\
ECC2K-358 & 358 & 227 & 31.7 & 17.6 & 60.3 & 31.1 & 127.9 & 65.5 & 131.9 & 66.9 \\
ECC2-353 & 353 & 103 & 6.4 & 4.3 & 6.4 & 4.3 & 108.9 & 55.5 & 158.3 & 80.2 \\
E-382  & 381  & 165 &  33.7 & 19.2  &  64.6  & 34.2  &  66.0 & 34.0     & 160.8&  82.2\\
Ed448 & 446 & 249  & 33.1 & 18.9 & 64.7 &  34.2 & 128.8 & 66.2 &  160.8 & 82.2\\
B-409  &409 & 124  & 29.1 & 15.6  &  63.4  & 33.3   &  129.1  & 66.7& 160.6 & 82.3 \\
BP384r1 & 384 & 206 & 33.3 & 18.7 & 66.0 & 35.5 & 130.0 & 67.6 & 160.7 & 82.3 \\
\noalign{\smallskip}\hline
\end{tabular}
\end{table*}

\begin{table*}[htb]
\caption{Curves providing security at least $256$ bits}
\label{tab:weak256}       
\begin{tabular}{lrrrrrrrrrr}
\hline\noalign{\smallskip}
Curve & $b(p)$ & $b(p_{m})$ & ${N_{2^{32}}}$ & ${C_{2^{32}}}$ &  ${N_{2^{64}}}$ & ${C_{2^{64}}}$ & ${N_{2^{128}}}$ & ${C_{2^{128}}} $ & ${N_{2^{160}}}$    &  ${C_{2^{160}}} $ \\
\noalign{\smallskip}\hline\noalign{\smallskip}
E-521 & 519 & 443 & 18.1 & 10.0 & 58.9 & 30.5 & 76.0 & 39.0 & 76.0 & 39.0 \\
P-521 & 521 & 391 & 31.4 & 16.7 & 50.0 & 26.0 & 128.8 & 66.3 & 130.5 & 66.2 \\
M-511   & 509   & 164 &  19.7  & 10.9  &  55.3 & 28.7   &  127.4  & 64.7  &  148.0& 75.5 \\
B-571  & 570 & 183 &  27.7  & 14.9  &  63.5 & 33.2   & 105.8  & 53.9  & 156.9  &  79.9\\
BP512r1 & 512 & 314 & 35.0 & 20.6 & 68.1 & 37.7 & 132.7 & 70.3 & 163.3 & 85.0 \\
K-571  & 570 & 161 &   36.1 & 21.7   &  67.1  & 36.9   & 131.8  & 69.5     &  164.8  & 86.5\\
\noalign{\smallskip}\hline
\end{tabular}
\end{table*}


\begin{table*}[htb]
\caption{Updated Pairing Friendly Curves providing $128$-bit security}
\label{tab:pairing128}       
\begin{tabular}{lrrrrrrrrrr}
\hline\noalign{\smallskip}
Curve & $b(p)$ & $b(p_{m})$ & ${N_{2^{32}}}$ & ${C_{2^{32}}}$ &  ${N_{2^{64}}}$ & ${C_{2^{64}}}$ & ${N_{2^{128}}}$ & ${C_{2^{128}}} $ & ${N_{2^{160}}}$    &  ${C_{2^{160}}} $ \\
\noalign{\smallskip}\hline\noalign{\smallskip}
KSS16\cite{barbulescu2019updating} & 263 & 131 & 31.2 & 17.1 & 66.5 & 36.1 & 129.2 & 67.0 & 158.8 & 80.7 \\
KSS18\cite{barbulescu2019updating} & 256 & 120 & 22.2 & 12.1 & 60.1 & 31.1 & 131.6 & 69.4 & 160.9 & 82.4 \\
BN\cite{barbulescu2019updating} \cite{cfrg2019pairing} & 462 & 289 & 35.8 & 21.3 & 67.4 & 37.1 & 131.4 & 69.2 & 162.9 & 84.4 \\
BLS12-381\cite{cfrg2019pairing} & 255 & 28 & 37.3 & 22.8 & 72.0 & 41.5 & 137.8 & 75.4 & 169.3 & 91.0 \\
BLS12\cite{barbulescu2019updating} & 308 & 73 & 37.3 & 23.0 & 71.9 & 41.6 & 137.6 & 75.3 & 169.6 & 91.4 \\

\noalign{\smallskip}\hline
\end{tabular}
\end{table*}

\begin{table*}[htb]
\caption{Updated Pairing Friendly Curves providing $192$-bit security}
\label{tab:pairing192}       
\begin{tabular}{lrrrrrrrrrr}
\hline\noalign{\smallskip}
Curve & $b(p)$ & $b(p_{m})$ & ${N_{2^{32}}}$ & ${C_{2^{32}}}$ &  ${N_{2^{64}}}$ & ${C_{2^{64}}}$ & ${N_{2^{128}}}$ & ${C_{2^{128}}} $ & ${N_{2^{160}}}$    &  ${C_{2^{160}}} $ \\
\noalign{\smallskip}\hline\noalign{\smallskip}
BLS24\cite{barbulescu2019updating} & 449 & 63 & 38.2 & 24.0 & 72.1 & 42.0 & 138.2 & 76.1 & 170.7 & 92.6 \\
KSS18\cite{barbulescu2019updating} & 502 & 87 & 41.6 & 27.4 & 77.7 & 47.5 & 145.3 & 83.2 & 178.2  & 100.0 \\
\noalign{\smallskip}\hline
\end{tabular}
\end{table*}

\begin{table*}[htb]
\caption{Updated Pairing Friendly Curves providing $256$-bit security}
\label{tab:pairing256}       
\begin{tabular}{lrrrrrrrrrr}
\hline\noalign{\smallskip}
Curve & $b(p)$ & $b(p_{m})$ & ${N_{2^{32}}}$ & ${C_{2^{32}}}$ &  ${N_{2^{64}}}$ & ${C_{2^{64}}}$ & ${N_{2^{128}}}$ & ${C_{2^{128}}} $ & ${N_{2^{160}}}$    &  ${C_{2^{160}}} $ \\
\noalign{\smallskip}\hline\noalign{\smallskip}
KSS36\cite{kiyomura2017secure} & 669 & 652 & 17.2 & 9.6 & 17.2 & 9.6 & 17.2 & 9.6 & 17.2 & 9.6 \\
KSS32\cite{kiyomura2017secure} & 738 & 591 & 33.9 & 19.2 & 65.8 & 35.3 & 129.2 & 66.8 & 146.6 & 74.3 \\
KSS18\cite{barbulescu2019updating} & 1108 & 341 & 30.4 & 16.2 & 67.9 & 37.5 & 132.8 & 70.5 & 163.4 & 85.1 \\
BLS42\cite{kiyomura2017secure} & 516 & 178 & 38.6 & 24.5 & 72.2 & 42.1 & 137.3 & 75.2 & 169.7 & 91.5 \\
BLS24\cite{kiyomura2017secure}  & 872 & 190 & 40.1 & 25.9 & 74.3 & 44.2 & 139.1 & 76.9 & 171.6 & 93.4 \\
BLS48\cite{cfrg2019pairing}\cite{kiyomura2017secure} & 518 & 91 & 39.9 & 25.7 & 74.8 & 44.6 & 140.7 & 78.6 & 173.1 & 95.0 \\
BLS24\cite{barbulescu2019updating} & 827 & 86 & 41.3 & 27.2 & 77.1 & 47.0 & 145.4 &  83.3 & 178.7  & 100.6 \\
\noalign{\smallskip}\hline
\end{tabular}
\end{table*}

Our data show that many curves have an abundance of weak keys at all levels, due to rather smooth factorizations 
of $p-1$ and, in particular, many divisors of $p-1$ of size $B$ and below.  The actual counts of weak keys vary, 
but roughly half of the curves surveyed have around $2^{160}$ weak keys at the level $B=2^{160}$.  There are also 
some notable examples of curves that have remarkably few weak keys, especially secp224k1, Brainpool256r1, and ECCp-359.
Some curves, such as secp193r2, Brainpool224r1, Curve25519, and ECC2-353 have very few weak keys at lower bounds but
many at $B=2^{160}$.  

There does not appear to be any bias of curves from one particular group or standard towards few or many weak keys.  
For example, curves from the NIST standard span the spectrum, with some curves such as P-224, P-192, P-384, and P-521 
having relatively few weak keys, but others such as K-163, P-256, K-283, B-409, and K-571 having relatively many.  The 
Brainpool standard includes some of the curves with the fewest weak keys as indicated above, but also has a few with 
relatively many such as Brainpool384r1 and Brainpool512r1.  The secp256k1 curve mentioned earlier, that is used in 
Ethereum and Bitcoin, falls roughly in the middle of the curves at the $128$-bit security level, and does not generally stand out 
in any way.
Except for KSS36, the pairing-friendly curves tend to have somewhat larger numbers of weak keys than the other curves, due in part to the fact that larger groups are required to compensate for the faster finite field discrete logarithm algorithms.  Of the three type of pairing-friendly curves, the BLS curves have the most weak keys at the 128- and 256-bit security levels, but the least at the 192-bit level.
All of this is consistent with the fact that current standards and practices place no restrictions on the
factorization of $p-1$, so we would expect its factorization to resemble that of a random integer.

\section{Application to the Certicom Challenge Curves}
\label{sec:certicom}

We have implemented in Sage \cite{sagemath} both the baby-step giant-step and Kangaroo algorithms from Section~\ref{sec:2} 
for testing whether a elliptic curve public-key comes from a weak private key.  We use the KKM \cite{KKM2007-remarks} 
extension for both algorithms, and use Python dictionaries for the required searchable lists so that searching is as efficient as possible.

To use the KKM method, as described at the end of Section~\ref{sec:2.1} 
we need to find a value of $c$ that minimizes the total number of group operations.  Recall that the total number of group 
operations, including computing the lookup table, is at most
\begin{equation*}
    c (\log_2 p + p^{1/c}) + 2 (c-1) \sqrt{d} \enspace.
\end{equation*}
for baby-step giant-step, and the same expression gives a reasonably accurate estimate for the kangaroo method.
In our implementation, given $p$ and $d$, we simply compute a local minimum of this expression.  We also set a hard constraint 
for the size of the lookup table at $2^{32}$ group elements, so if the value of $c$ obtained via minimization causes the table
size $c p^{1/c}$ to be too large, we increased $c$ until the table size was below this bound.

The additional functions and parameters used in the implicit kangaroo algorithm are as follows:
\begin{itemize}
    \item \textbf{Partition function}: The elliptic curve group $E$ is partitioned into $L$ parts as 
    $E = S_1  \cup  S_2  \cup  ... \cup  S_{L} $ via a partition function $f: E \rightarrow \{1, 2, ..., L\}$.
    We use $L=2^n$, and for a point $P \in E$ we set $f(P)=j$ if the $n$ least-significant bits of the integer 
    representation of the $x$-coordinate of $P$ is $j-1$.  We used $L=1024$ in our implementation.
    
    \item \textbf{Jump set}: For each partition $S_j$, a small kangaroo jump $s_j$ is selected as a random
    integer between $1$ to $\sqrt{d}/2$ with the aim of maintaining the mean step size of the jumps close to 
    $\sqrt{d}/2$, where $d$ is the order of the subgroup $D$ of $\mathbb{F}_p^*$ in which the discrete logarithm
    is assumed to lie.  The first $L-1$ jumps are selected randomly, and the last is selected to ensure that the
    mean step size condition is satisfied.
    
    \item \textbf{Distinguished point property}: Distinguished points help to detect collision between the
    elliptic curve points. For a fixed positive integer $t$, we call a rational point $P \in E$ a distinguished
    point of the curve if $t$ least significant bits of integer representation of its $x$-coordinate are $0$.
    Since the integer $t$ determines the proportion of distinguished points in the elliptic curve, the integer
    $t$ can be used to control the storage overhead of the algorithm.
\end{itemize}
Both implementations were tested using one safe curve M221 and one Certicom curve ECCp131. 
1000 random instances of weak discrete logs were generated, and the code for baby-step giant-step successfully found the weak key
in all cases.  The kangaroo algorithm finds the weak key with success rate of more than $95$ percent, but the failed cases
can also be solved in a few more trials if we take different random jump set. 

As an application, we used our baby-step giant-step implementation to verify that each of the 14 remaining unsolved 
Certicom Challenge \cite{certicom2009challenge} elliptic curve discrete logarithm problem instances is such that the
discrete logarithms are not in any subgroup of $\mathbb{F}_p^*$ of order less than the bound $B = 2^{48}$.  
The Certicom Challenges 
were put forward by Certicom in 1997 in order to stimulate research into the elliptic curve discrete logarithm problem.
Nine problem instances over characteristic two finite fields (including four Koblitz curves), and five instances over
prime finite fields remain unsolved.  The field sizes range from $131$ to $359$ bits.

Because our expectation was that we would only be able to certify that the discrete logarithms are not weak, rather than 
actually computing any, we also generated and solved a random discrete logarithm instance that was weak for each curve 
as an extra test of our implementation.  In all cases the random discrete logarithm was successfully recovered.

The data related to our computations is given in Table~\ref{tab:certicom}. For each curve, we list the number of subgroups 
that had to be tested according to the method of Section~\ref{sec:weaktest}, the size (in bits) of the largest subgroup order $b(d_{max})$,
and the total CPU time in minutes for the test.  The breakdown of time spent in the baby step and the giant step stages is 
not listed, as these each required approximately half of the total time.  Each stage performs the same number ($\sqrt{d}$) 
of group operations for testing a subgroup of size $d$.  The giant steps were slightly faster, due to the fact that during 
the baby step stage the list of baby steps is created, and that is slightly more expensive than the searching done in the giant step stage.
The computations were run on a shared memory machine with 64 Intel(R) Xeon(R) X7560 running at 2.27 GHz and 256 GB of RAM running Linux.

\begin{table}
 \caption{Certicom Challenge Data}
 \label{tab:certicom}       
 \begin{tabular}{lrrrrr}
 \hline\noalign{\smallskip}
Curve & \# Subgroups & $b(d_{max})$ & Total Time \\
\noalign{\smallskip}\hline\noalign{\smallskip}
ECC2-131 &   1 & 18 &          $0.05$ m \\
ECC2-163 & 186 & 48 & 57d  7h $34.46$ m \\
ECC2-191 &   3 & 48 &     11h $22.98$ m \\
ECC2-238 &   3 & 46 &      7h $54.27$ m \\
ECC2-353 &   1 &  7 &          $0.03$ m \\
\noalign{\smallskip}\hline
ECC2K-130 &  7 & 48 & 1d  8h $35.92$ m \\
ECC2K-163 &  4 & 48 &    19h $46.66$ m \\
ECC2K-238 & 12 & 48 & 3d 10h $12.92$ m \\
ECC2K-358 &  5 & 31 & 2d 16h $43.40$ m \\  
\noalign{\smallskip}\hline
ECCp131 & 2 & 48 & 5h $16.19$ m \\
ECCp163 & 1 & 29 &     $0.95$ m \\ 
ECCp191 & 1 & 48 & 7h $27.08$ m \\
ECCp239 & 1 & 23 &     $0.27$ m \\
ECCp359 & 1 &  6 &     $0.02$ m \\
\noalign{\smallskip}\hline
\end{tabular}
\end{table}

There is some variability in the number of subgroups required to test.  In most cases the number is quite small, 
most often 1, but in the case of ECC2-163, $186$ different subgroups had to be 
tested, with the majority of these being roughly of size $2^{48}$ and requiring more than 6 hours each; this curve required by far the most computational effort, close to two months.  On the other hand, many of the verifications finished very quickly,
most notably ECC2-353 and ECCp359 which have only one subgroup of order $82$ and $36$ less than $2^{48}$, 
respectively.

We note that we did not manage to solve any of the Certicom challenges with this approach.  This is of course completely
as expected, as the probability that a randomly-selected private key would be weak is very small, and there would have
been no reason for Certicom to generate the private keys used for the challenges in any other way.

\section{Conclusions and Future Work}
\label{sec:6}

The weak keys described in this paper are somewhat more subtle than those found in 
Ethereum \cite{ethercombing}, 
in the sense that they do not show any \textit{obvious weaknesses} on the surface, such as being a very small
integer.  Rather, they look the same as a typical random integer generated as per standard practices
and guidelines such as those in \cite[Sec.~5.6.1.2]{NIST-DLkeys} in terms of expected bit length; their 
weakness comes from the underlying multiplicative structure of $\mathbb{F}_p^*$ via implicit group 
representations.  The Ethereum weak keys can also be made to look like random integers by selecting 
them from any short interval known to the attacker.  Such weak keys have the same properties as ours; 
both look to be random integers, can be easily generated, and can be easily recovered 
from the public key.  The main difference is that keys from a short interval \textit{cannot be prevented} other
than ensuring that keys are generated verifiably at random, whereas ours can be \textit{eliminated} systematically 
by ensuring that the group order $p$ is a safe prime.  

Our main conclusion is therefore that the prime group order for discrete logarithm based cryptosystems 
should be subject to similar considerations as the prime divisors of RSA moduli.  Weak RSA keys
(RSA moduli that can be factored easily) can be avoided by ensuring that the prime divisors are safe,
and our type of weak keys can be avoided by using groups whose prime order is safe.  The current best
practice for RSA moduli, generating verifiably at random, also applies to discrete logarithm based 
cryptosystems as our type of weak keys do not occur with very high probability as long as keys are 
generated randomly.  Thus, random key generation is also sufficient to avoid our type of weak keys, 
provided that key generation software is audited to ensure that it does in fact generate keys randomly.
In the case of discrete logarithm based systems, where a single group is provided for 
multiple users (for example, standardized elliptic curves) as opposed to one per user as in RSA, 
taking the extra step to ensure that safe prime group orders are used is a feasible option that would
either completely eliminate or at least reduce the ability of malicious key generators generating 
our type of weak keys.

One direction of further work is to repeat the experiments of \cite{ethercombing} to test Ethereum keys,
but instead of searching for small numerical private keys to search for the type of weak keys described 
in this paper in an attempt to identify more vulnerable Ethereum and Bitcoin wallets than those already given in Tables~\ref{tab:bitcoin} and \ref{tab:ethereum}. 
Similarly, it would be interesting to expand the search for weak keys in SSH, 
TLS, and the Australian e-ID card implementations, conducted in \cite{BHHMNW14}, to include the type of 
weak keys described here. Current recommendations for private key 
generation, including \cite[Sec.~5.6.1.2]{NIST-DLkeys}, only demand that the key be selected 
uniformly at random between $1$ and $p-1$, and investigations such as that in \cite{BHHMNW14} consider 
only this property. If keys are indeed selected randomly then we would not expect to be able to recover 
any using our methods because the probability is too low.  On the other hand, finding weak keys could be
evidence that this vulnerability had been previously discovered and secretly exploited to serve as a trap
door for an adversary.

Searching for additional types of weak keys is also an intriguing possibility.  For example, in his first 
work on the subject, Cheon \cite{cheon,cheon-JCrypt} also describes how auxiliary inputs can be 
used to compute discrete logarithms in a group of order $p$ when $p+1$ has small divisors.  Subsequent 
work by Satoh \cite{Satoh2009-generalization}, for example, has generalized this to the case where the 
cyclotomic polynomial $\Phi_n(p) = p^{n-1} + p^{n-2} + \dots + 1$ has small divisors (note that $\Phi_2(p) 
= p+1$).  The main idea behind these extensions is to first embed $\alpha$ into an element of the order 
$\Phi_n(p)$ subgroup of $\mathbb{F}_{p^n}^*$, and then use the oracle to lift the problem to a smaller 
subgroup.  In principal, one could hope that after embedding $\alpha$ into the order $\Phi_n(p)$ subgroup 
that it already lands in the smaller subgroup, and then declare the corresponding key $g^\alpha$ to be 
weak.  However, another feature of these extensions is that they generally require the implicit evaluation 
of $g$ to the power of a polynomial in $\alpha,$ thereby requiring more auxiliary inputs 
to obtain $g^{\alpha^i}$ for various values of $i$; we were unable to find any families of keys for which 
these types of generalizations can be applied without requiring auxiliary inputs.

Another interesting question is what other discrete logarithm algorithms can be realized using implicit 
group representations.  Of the generic algorithms, we have seen that both baby-step giant-step and Pollard 
kangaroo work.  Pohlig-Hellman using the implicit group representation would be especially interesting 
since so many of the recommended elliptic curves are such that $p-1$ is relatively smooth.  However, that 
would require being able to lift $g^{\alpha}$ into each prime-power order subgroup of $\mathbb{F}_p^{\times}$.  
It is not clear how that can be done, other than appealing to a Diffie-Hellman oracle or auxiliary inputs as Brown and Gallant 
\cite{brown2004static} and Cheon \cite{cheon,cheon-JCrypt} do.   Using index calculus to attack the
discrete logarithm problem in $\mathbb{F}_p^*$ using implicit representations in an effort to avoid the
need for nice smoothness properties in the group $G$ itself also appear to be fruitless.

All the ideas and applications of implicit group representations use the fact that exponentiation 
in $G$ is a group action of the order $p-1$ multiplicative group $\mathbb{F}_p^*$ on $G$.  Consequently, 
as observed by Smith~\cite[p.23]{Smith2018-isogenies}, analogues of the work of Brown and
Gallant~\cite{brown2004static}, Cheon~\cite{cheon,cheon-JCrypt}, and others to cryptosystems based on 
other types of group actions, including isogengy-based cryptography, are possible.  We observe that 
similar types of weak keys as described here are will therefore also exist in those settings.




\begin{acknowledgements}
The first author is supported in part by NSERC of Canada. The second author is supported by MeitY, India under the ISEA (Phase II) project.
\end{acknowledgements}

%

\bibliographystyle{spmpsci}      
\bibliography{kangaroo.bib}   

\end{document}